\documentclass[
aps,
prl,
reprint,
superscriptaddress
]{revtex4-2}
\usepackage[normalem]{ulem}
\usepackage{color}
\usepackage{amsmath,amssymb}
\usepackage{graphicx}
\usepackage{hyperref}

\begin{document}

\title{Topological Word for Non-Abelian Topological Insulators}

 \author{Zhenming Zhang}
 \affiliation{Laboratory of Quantum Information, University of Science and Technology of China, Hefei 230026, China}
 \author{Tianyu Li}
\email{tianyuli@m.scnu.edu.cn}
\affiliation {Key Laboratory of Atomic and Subatomic Structure and Quantum Control (Ministry of Education), Guangdong Basic Research Center of Excellence for Structure and Fundamental Interactions of Matter, School of Physics, South China Normal University, Guangzhou 510006, China}
\affiliation {Guangdong Provincial Key Laboratory of Quantum Engineering and Quantum Materials, Guangdong-Hong Kong Joint Laboratory of Quantum Matter, Frontier Research Institute for Physics, South China Normal University, Guangzhou  510006, China}
\author{Wei Yi}
\email{wyiz@ustc.edu.cn}
\affiliation{Laboratory of Quantum Information, University of Science and Technology of China, Hefei 230026, China}
\affiliation{Anhui Province Key Laboratory of Quantum Network, University of Science and Technology of China, Hefei 230026, China}
\affiliation{CAS Center For Excellence in Quantum Information and Quantum Physics, Hefei 230026, China}
\affiliation{Hefei National Laboratory, University of Science and Technology of China, Hefei 230088, China}

\date{\today}

\begin{abstract}
We propose a unified framework, dubbed topological word, for the complete non-Abelian bulk-boundary correspondence in multigap non-Abelian topological insulators.
Composed by an ordered sequence of letters, each a non-Abelian charge depicting the gap-resolved topology, the topological word captures both the global non-Abelian topology corresponding to the homotopy classification, and the band-adjacency information. The latter, though crucial for the edge-state pattern across multiple gaps, is often overlooked in previous studies.
We confirm our framework using both static models and periodically driven Floquet systems,
and discuss its connection and distinction with existing descriptions, such as the phase-band singularities and braiding representations.
Intriguingly, topological word continues to provide insight regarding topology and edge states, even as the global non-Abelian topology becomes ill-defined under broken parity-time symmetry.
\end{abstract}

\maketitle

{\it Introduction.---}
Topological band theory provides a powerful framework for understanding robust boundary phenomena in condensed-matter systems, where bulk topological invariants depict the band topology according to the homotopy classifications, and account for robust edge states through the bulk–boundary correspondence (BBC)~\cite{QiZhang2011, Shenbook2012, HasanKane2010, AltlandZirnbauer1997, Ryu2010, Chiu2016, Graf2013, Kitaev2009, BernevigHughes2013, Asboth2016}. 
In contrast to prevalent examples involving only a single band gap and characterized by Abelian invariants~\cite{QiZhang2011, Ryu2010, AltlandZirnbauer1997}, recent studies have revealed a general class of multigap topological insulators~\cite{Wu2019, Qiu2023, Guo2021, Jiang2021, Wang2024, Tiwari2020, Li2023, Yang2024, LinPXue2026, Jiang2021NP,YHu2024,Jankowski2024, HQiu2026,Jiang2022,Jiang2026,Yang2019,Jiang2024,Neup2021,Slagerprb2024}, where the tangled geometry of eigenstates necessitates the description of non-Abelian topological charges~\cite{Wu2019, Yang2019, Guo2021} or braiding processes~\cite{GCMa2025}.

\begin{figure}[tbp]
    \centering
    \includegraphics[width=\linewidth]{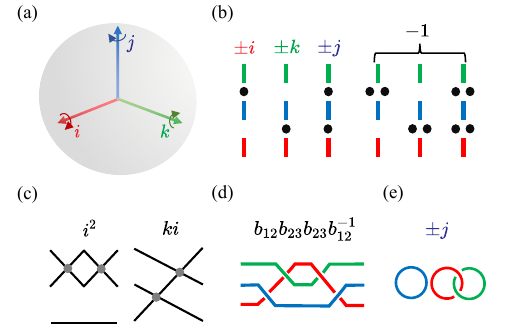}
    \caption{
    Illustration of the $\mathbb{Q} _8$ topology and the topological word in a three-band system.
    (a) Quaternion charges $ i, j, k$ correspond to $\pi$ rotations with respect to different axes of an orthonormal eigenframe, as the Brillouin zone is traversed. For instance, charge $i$ indicates a $\pi$ rotation of the eigenstates in the upper two bands (blue and green) relative to that of the lowest band (red). As a consequence, the upper two bands (or the upper gap) acquire a nontrivial $\mathbb{Z}_2$ topology.
    (b) Typical relations between quaternion charges and edge-state (black dots) configurations in a static system.
    Colored lines denote the energy bands, in accordance with the colored eigenframe in (a).
    (c) Correspondence between topological words and Dirac singularities, illustrated for $i^2$ and $ki$, respectively.
    (d) Braid word of the charge $Q = j$ with the corresponding braiding process.
    (e) A three-component linking structure corresponding to quaternion charges $\pm j$. 
    }
    \label{fig1}
\end{figure}

An outstanding example is the parity-time (PT) symmetric three-band topological insulators~\cite{Wu2019,Guo2021}, whose Bloch eigenstates are real and form an orthonormal frame in $\mathbb{R}^3$. The global topology of this frame over the Brillouin zone is classified by the quaternion group $\pi_1(M_3)=\mathbb{Q} _8$, where $M_3=O(3)/O(1)^3$, and $O(N)$ is the orthogonal group in $N$ dimensions. Here the quaternion group $\mathbb{Q} _8= \{\pm1,\pm i,\pm j,\pm k\}$, with the quaternions satisfying $i^2=j^2=k^2=ijk=-1$, and the anti-commutative relations between each pair (for instance $ij=-ji$)~\cite{HatcherTopoBook}.  
Geometrically, the group elements correspond to rotations of the eigenframe over the Brillouin Zone [see Fig.~\ref{fig1}(a)].
Such a non-Abelian band topology has attracted much attention of late, but despite the clarity of the bulk homotopy classification, the BBC remains incomplete.
For instance, systems with the same quaternion charge $Q=-1$ can exhibit distinct edge-state patterns [Fig.~\ref{fig1}(b)]~\cite{Guo2021}. 
Similar issues also arise in periodically driven Floquet systems, where the presence of anomalous edge states within the quasienergy $\pi$ gap further complicates matter~\cite{Li2023, LinPXue2026, HQiu2026}. Though the BBC can be restored therein using phase-band singularities in the synthetic momentum-time manifold~\cite{Li2023}, the scheme fails to apply in static systems.

In this work, we show that a genuine non-Abelian BBC can be achieved by introducing \emph{topological words}, which consistently apply in both the static and Floquet systems, thus offering a unified description.
A crucial observation is that the homotopy classification retains only the global topology of the eigenframe, whereas the band-adjacency information, or equivalently the relative positions of the multiple gaps, also plays an important role in determining the edge-state patterns.
The topological word, which consists of a sequence of quaternions indicating gap-resolved topology, retains both the global homotopy and band-adjacency information, thereby capable of achieving the non-Abelian BBC.
We illustrate how these topological words can be constructed from Dirac points within the continuous manifold interpolating bulks with different parameters [Fig.~\ref{fig1}(c)], a strategy extendable to Floquet non-Abelian topological insulators.
Topological words also connect naturally to braid representations [Fig.~\ref{fig1}(d)] and Hopf links [Fig.~\ref{fig1}(e)] of multiband eigenstates, and remain informative even beyond the PT-symmetric regime.

{\it Topological word.---}
Let us first examine systems with their global topology characterized by $i$, $j$, or $k$. From the perspective of homotopy classification, these non-Abelian charges appear to have cyclic symmetry, all depicting a $\pi$ rotation of the eigenframe [Fig.~\ref{fig1}(a)]. 
However, when the overall topological charge is $i$ or $k$, the system hosts a single pair of edge states, whereas for charge $j$, a pair of edge states appears in each gap [Fig.~\ref{fig1}(b)]. This is because the rotation corresponding to the charge $j$ involves eigenstates of the first and third bands (red and green), which are not adjacent, so that both gaps acquire nontrivial Zak phases and edge states~\cite{Wu2019, Guo2021, Slager2024}. This example shows that the distribution of edge states depends crucially on the band-adjacency information, which is lost in the homotopy classification where only the overall quaternion charge matters.

Motivated by this observation, we introduce a topological word description, where the overall non-Abelian charge is expressed as an ordered product
\begin{equation}
Q=Q_1 Q_2 \cdots Q_n,\qquad Q_i\in\mathcal{E}.\label{eq:word}
\end{equation}
Here the sequence $Q_1 Q_2 \cdots Q_n$ is defined as a topological word, and
$\mathcal{E}$ denotes the set of letters, each corresponding to the $\mathbb{Z}_2$ topology of a pair of adjacent bands (or equivalently of the intervening gap). 
Although each letter indicates the Abelian topology of a single gap, the multigap scenario renders the letters non-commutative.
While the overall topological charge $Q$ is obtained by performing the product according to Eq.~(\ref{eq:word}), the sequence of letters in a word records the gap-resolved topology. 
The key idea behind this construction is that all topological invariants can be decomposed into elementary contributions associated with adjacent-band topology, which serve as the generators of the homotopy group.

Here two remarks are in order. First, for a given gap, the number of letters it contributes to the word equals the number of edge-state pairs appearing in that gap, which is bounded, for instance, by the maximum hopping range of the lattice model. Second, the word representation also contains a gauge redundancy: letters can be locally rewritten using the algebra of the underlying non-Abelian group. Different words related by this gauge transformation predict the same edge-state patterns~\cite{supp}.

Take, for example, a three-band topological insulator with only nearest-neighbor and next-nearest-neighbor hoppings. Here the allowed letter set $\mathcal{E}=\{\pm i,\pm k\}$, where
$\pm i$ and $\pm k$ correspond to nontrivial $\mathbb{Z}_2$ topology of the upper and lower gap, respectively. 
A trivial topology contributes no letters and is therefore identified as the empty word $1$.
Due to the hopping-range limitation, the maximum number of letters allowed for each gap is $2$, and words such as $k^3$ are forbidden. 
Consequently, a total charge $Q=k$ admits only the word $k$, corresponding to a single pair of edge states in the lower gap. The case with $Q=i$ is similar, but with the edge-state pair in the upper gap.
The phase with $Q=j$ is given by the word $ki$ (or equivalently $-ik$ under a gauge transformation), which by construction indicates an edge-state pair in each gap. 
Importantly, when $Q=-1$, three distinct words are allowed: $k^2$, $i^2$, and $kiki$, which respectively correspond to the three edge-state patterns shown in Fig.~\ref{fig1}(b). 
When the hopping range becomes longer, more complicated words become possible, and the total quaternion charge $Q$ alone becomes even less predictive of the edge-state patterns, as exemplified by the $Q=-1$ case above~\cite{supp}.

Topological word is also applicable to Floquet systems, where the quasienergy spectrum is periodic~\cite{Oka2019}. The lowest and highest bands are then also adjacent, which, for a three-band system, enlarges the allowed letter set to $\mathcal{E}=\{\pm i,\pm j,\pm k\}$. As a consequence, a trivial overall charge $Q=1$ may arise not only from an empty word $1$, but also from nontrivial combinations such as $-ijk$. The latter corresponds to the recently observed anomalous edge states in Floquet non-Abelian topological insulators~\cite{Li2023, LinPXue2026,HQiu2026}.

{\it Topological Word from Dirac points.---}
We now address two questions: how to extract the topological word from a given Hamiltonian and why it works. 
These can be addressed by considering Dirac singularities in the continuous manifold interpolating topologically distinct bulks.

\begin{figure}[tbp]
    \centering
    \includegraphics[width=\linewidth]{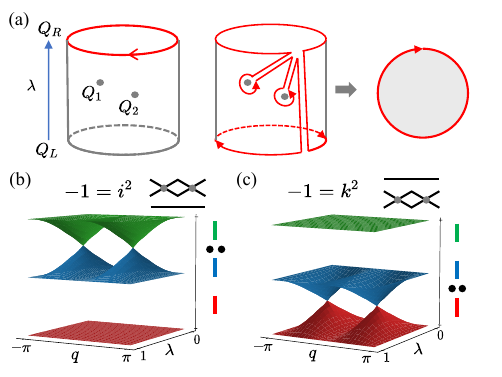}
    \caption{
    Singularities and their relation to the topological word.
    (a) Illustration of an interpolation between two phases carrying charges $Q_L$ and $Q_R$, with singularities in between.
    The Brillouin zone and the interpolation parameter $\lambda$ form a cylinder $S^{1}\!\times[0,1]$.
    Introducing branch cuts from a base point allows the loops around the singularities and the boundary loops to be combined into a composite loop that can be continuously deformed to the boundary of a simply connected disk.
    (b)(c) Transitions between the trivial phase and phases characterized by $i^2$ and $k^2$, respectively. The bare band crossings are schematically represented in the upper right corner.
    }
    \label{fig2}
\end{figure}

To interpolate between two bulks governed by Hamiltonians $H_L$ and $H_R$, we introduce an interpolation $H(\lambda)$, with an auxiliary parameter $\lambda\in[0,1]$ such that $H(\lambda = 0)=H_L$ and $H(\lambda = 1)=H_R$. 
Thus, the Bloch Hamiltonian $H(\lambda,q)$ is defined on a cylindrical manifold
$S^{1}\times[0,1]$, spanned by the quasimomenta $q$ in the Brillouin zone and the interpolation parameter $\lambda$ [Fig.~\ref{fig2}(a) left].
The topological charge is obtained from a loop integral along a closed path within this manifold~\cite{supp}.
Integrating along the upper rim of the cylinder [Fig.~\ref{fig2}(a) left] yields the topological charge $Q_R$ of $H_R$, while integrating over the lower rim gives $Q_L$ of $H_L$. If the two bulks are topologically distinct, singularities must emerge in the manifold. 
The topological charge of a given singularity can be evaluated through a loop integral encircling it, denoted as $Q_{n}$ ($n=1,2,\cdots$).
To connect these charges with $Q_R$ and $Q_L$, we choose a base point on the cylinder and introduce branch cuts connecting these loops. The resulting loops can then be combined into a composite one that encloses no singularities [Fig.~\ref{fig2}(a) middle]. Different choices of the base point and branch cuts amount to different gauges and have no physical consequences~\cite{supp}. This composite loop can be continuously deformed into the boundary of a simply connected disk on the cylinder [Fig.~\ref{fig2}(a) right], whose boundary therefore carries trivial topology~\cite{HatcherTopoBook}. Hence, we obtain 
 \begin{equation}
Q_R^{-1}(Q_1 Q_2\cdots)Q_L=1,
\end{equation}
where the order of product is dependent on the choice of gauge. 
We then have the domain-wall charge
\begin{equation}
\Delta Q:=Q_R/Q_L=Q_1 Q_2\cdots.
\end{equation}
Note that the interpolation $H(\lambda,q)$ need not be unique, which in general leads to different sequences of $\{Q_1,Q_2,... \}$. However, under smooth deformations of the interpolation, there always exists a choice of $H(\lambda,q)$ with the minimum number of singularities, which can be regarded as the canonical choice, with the resulting sequence the topological word.
With this canonical choice, each remaining Dirac point corresponds to a local mass inversion and gives rise to a domain-wall state~\cite{Shenbook2012}. The configuration of domain-wall states is thus encoded into the topological word.
For the open-boundary condition, we simply set $Q_L=1$, and the topological word so constructed would be associated with the bulk $H_R$.

As a concrete example, we examine the three-band tight-binding model in Ref.~\cite{Guo2021}, with  both nearest-neighbor and next-nearest-neighbor intercell hoppings
\begin{align}\label{eq:Ham}
&H(\lambda)=\sum_{n}\sum_{\alpha} s_{\alpha\alpha}\,c^\dagger_{\alpha,n}c_{\alpha,n}\nonumber \\
&+\sum_{n}\sum_{\alpha,\beta}\Big(
v_{\alpha\beta}c^\dagger_{\alpha,n}c_{\beta,n+1}
+v'_{\alpha\beta}c^\dagger_{\alpha,n}c_{\beta,n+2}
+\text{H.c.}\Big).
\end{align}
Here $\alpha,\beta\in\{A,B,C\}$ label the sublattice sites. The parameters $s_{\alpha\alpha}$, $v_{\alpha\beta}$, and $v'_{\alpha\beta}$ are all functions of $\lambda$, which interpolate between different Hamiltonians. Their detailed forms are given in the Supplemental Material~\cite{supp}.

In Fig.~\ref{fig2}(b) and \ref{fig2}(c), we numerically show two distinct interpolations, connecting a common trivial phase to different nontrivial phases with the same overall quaternion charge $Q=-1$.
As illustrated in Fig.~\ref{fig2}(b)(c), pairs of Dirac points appear in different gaps, which are characterized by the topological words $i^2$ and $k^2$, respectively, consistent with their distinct edge-state patterns.
Note that for clarity, we retain only the barebone band crossings, and represent the Dirac singularities with schematics similar to those in Fig.~\ref{fig1}(c).

The above construction can be naturally extended to Floquet systems, where the interpolation parameter $\lambda$ is identified with time $t\in [0,T]$. It follows that the topologically trivial Floquet operator $U(t=0)=1$ is continuously deformed into $U(t=T)$, and the resulting momentum-time manifold forms the so-called phase band~\cite{Nathan2015, Hu2020PRL, HuZhao2020}. The patterns of the phase-band singularities can be translated to topological words following the recipe above, which uniquely determines the edge-state configuration~\cite{Li2023, LinPXue2026}. 

\begin{figure}[tbp]
    \centering
    \includegraphics[width=\linewidth]{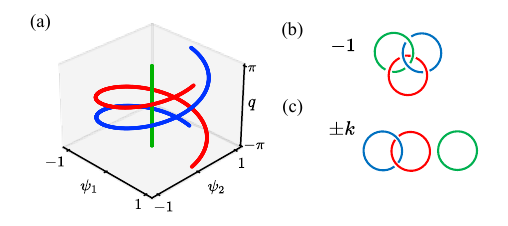}
    \caption{
    (a) Components $(\psi_1,\psi_2)$ of the three eigenstates as functions of the quasimomentum $q$, for parameters of the $k^2$ phase.
    (b) Linking structure for the topological charge $Q=-1$, where the three trajectories are pairwise Hopf  linked.
    (c) Linking structure for $Q=k$, where only two rings are linked.
    }
    \label{fig3}
\end{figure}

\begin{table*}[tbp]
    \centering
    \includegraphics[width=\linewidth]{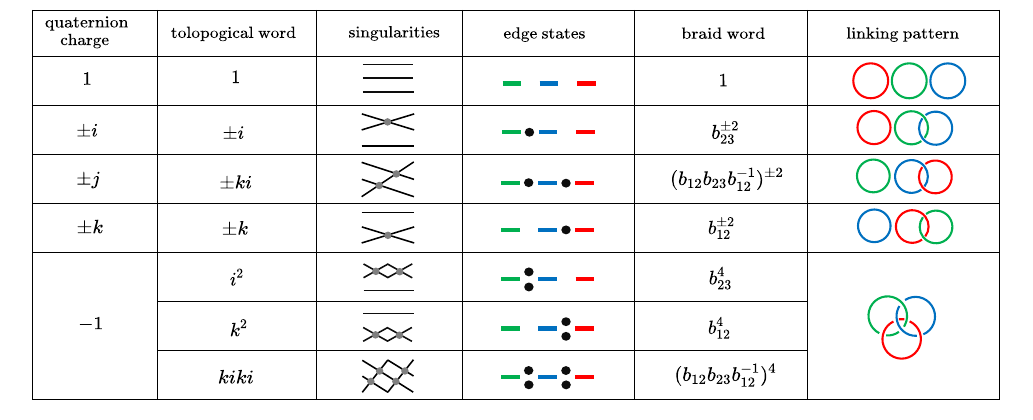}
    \caption{
    Summary of the correspondence between different descriptions of the non-Abelian topology
    for three-band models with nearest-neighbor and next-nearest-neighbor hoppings.
    }
    \label{tab1}
\end{table*}

{\it Connection with braids and Hopf links.---}
In a recent study~\cite{GCMa2025}, the rotation of the three-dimensional eigenframe across the Brillouin zone is mapped to the braiding process of three strands. 
Elementary braidings are generated by nearest-neighbor exchanges $b_{i,i+1}$, denoting the counterclockwise exchange of the $i$th and $(i+1)$th strands. Identifying the band adjacency with 
that of the strands leads to a correspondence between the gap-resolved topology with the so-called braid word. For example, the $i$ and $k$ charges correspond to the braid words $b_{23}^2$ and $b_{12}^2$, respectively. By contrast, the $j$ charge corresponds to the exchange of the first and third strands twice,
formally $b_{13}^2$. But since the two strands are not adjacent, the braid word must be written as $b_{12}b_{23}^2b_{12}^{-1}$ [Fig.~\ref{fig1}(d)], involving four elementary braidings~\cite{GCMa2025}. 
Thus, by keeping strand-adjacency information, properly constructed braid words can be translated to topological words, and correctly predict the edge-state patterns. 
However, edge-state patterns cannot be straightforwardly read off a braid word, unlike the case with topological words. And the translation from a braid word to topological word is often cumbersome.

Alternatively, at the cost of the band-adjacency structure, the braiding description can be visualized in terms of the intertwining of eigenstates.
In a three-band system, each eigenstate can be viewed as a three-component vector $(\psi_1,\psi_2,\psi_3)$ in a local eigenframe. 
While the homotopy classification corresponds to the rotation of the eigenframe, one can always choose a reference direction as the rotation axis. 
Taking the direction associated with $\psi_3$ as the rotation axis, for instance, one can track the trajectory of $(\psi_1,\psi_2)$ as the Brillouin zone is traversed.
For the overall topological charge $Q=-1$, the trajectories of all three eigenstates exhibit the winding pattern in Fig.~\ref{fig3}(a). With the $S^1$ nature of the Brillouin zone, the trajectories close into three rings whose linking structure reflects the overall non-Abelian topology. In particular, the $Q=-1$ case yields a structure with pairwise Hopf links [Fig.~\ref{fig3}(b)], whereas for $Q=\pm i,\pm j,\pm k$, only two rings are linked [Fig.~\ref{fig3}(c)]. Apparently, the linking structure, as the overall non-Abelian charge, is not sufficient for the non-Abelian BBC.

We summarize in Table~\ref{tab1} the relations among different descriptions of the non-Abelian topology. 
The table applies to static systems with nearest-neighbor and next-nearest-neighbor hoppings, whereas longer-range hoppings and Floquet systems allow more complicated words and richer edge-state patterns~\cite{supp}. 
Importantly, topological words uniquely determine the edge-state patterns in a straightforward fashion.

\begin{figure}[tbp]
    \centering
    \includegraphics[width=\linewidth]{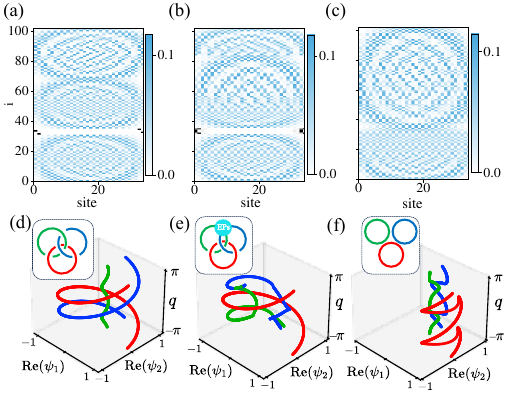}
    \caption{
(a)(c) Spatial distribution of eigenstates under the open boundary condition for $\gamma=0.3,\,0.6,$ and $1.1$, respectively.
The horizontal axis denotes the lattice site, and the vertical axis labels eigenstates, ordered by the real part of the eigenspectrum.
The blue color bar represents the probability density, and the edge modes are marked in black.
(d)(f) Trajectories of $(\text{Re}(\psi_1),\text{Re}(\psi_2))$ across the Brillouin zone for $\gamma=0.3,\,0.6,$ and $1.1$, respectively.
Red, blue, and green denote the three bands, ordered by the real part of the eigenspectrum.
Insets show the corresponding linking structrues.
}
    \label{fig4}
\end{figure}

{\it Beyond the non-Abelian topology.---}
The descriptive power of topological words persists even beyond the reign of non-Abelian topology. 
As an illustrative example, we show how topological words offer crucial information on the breakdown of non-Abelian topology and fate of edge states when the PT-symmetry is broken.

For this purpose, we introduce non-Hermitian perturbations to the three-band model (\ref{eq:Ham}), in terms of the non-reciprocal hoppings
\begin{equation}
H_{\text{nh}}
= \gamma \sum_{n}\left(
c^\dagger_{B,n} c_{C,n}
- \text{H.c.}
\right),
\end{equation}
where $\gamma$ denotes the non-Hermitian strength. 
We focus on the case with a global topological charge $Q=-1$, where the Abelian $\mathbb{Z}_2$ topology between any two adjacent bands are trivial.

For a small $\gamma$, the Bloch Hamiltonian remains PT symmetric, ensuring a real spectrum and real eigenstates, and the non-Abelian topology remains well-defined by invoking the biorthogonal framework~\cite{supp, biorthogonal, Ashida2020}.
Under open boundary conditions, two pairs of edge states emerge in the lower gap [Fig.~\ref{fig4}(a)], consistent with the topological word $k^2$, and the global non-Abelian topology can be visualized through the linking structure as before [Fig.~~\ref{fig4}(d)].

As $\gamma$ increases, the PT symmetry is eventually broken: the upper gap in the real spectrum closes, and the two upper bands become complex, with exceptional points (EPs) emerging in the momentum space~\cite{Ashida2020}. The quaternion charge is therefore ill-defined. 
Remarkably, however, the edge states persist [Fig.~\ref{fig4}(b)]. This can be understood from the topological word $k^2$ in the PT-unbroken regime, which suggests that the lower gap is essential for the band topology.
Although the upper two bands (blue and green) cease to be real and develop EP degeneracies, the lower gap (though complex now) remains open, suggesting remnant topology. Indeed, the eigenstate trajectory of the lowest band (red) winds around those of the upper bands [Fig.~\ref{fig4}(e)], and the resulting linking structure protects the surviving edge states. 
The non-Abelian topology is effectively reduced to an Abelian one associated with the remaining gap, and the topological word tracks the surviving topology and edge states even as the system undergoes PT-symmetry breaking.
When $\gamma$ increases further, the EPs disappear: a purely imaginary gap emerges between the upper two bands, and the lower gap closes and reopens, indicative of a phase transition.
Thereafter, the lowest band no longer winds around the upper two [Fig.~\ref{fig4}(f)], the system becomes topologically trivial, and the edge states disappear [Fig.~\ref{fig4}(c)]. Note that we expect topological words provide similar insights under Hermitian perturbations that break the PT symmetry.

{\it Discussion.---}
We have introduced topological words as a unified framework for the complete non-Abelian
BBC in multigap systems. The topological word naturally encodes both the global non-Abelian topology, and the gap-resolved information, thereby bridging conventional single-gap Abelian invariants and multigap non-Abelian topology. 
Since our construction is quite general, we expect topological words to apply in higher-dimensional topological insulators and multi-band systems with four or more bands, where the letters are no longer restricted to quaternions.

\acknowledgments
This work is supported by the National Natural Science Foundation of China (Grant Nos. 12374479, 12504189), and Quantum Science and Technology-National Science and Technology Major Project (Grant No. 2021ZD0301904). T.L. is supported by the Guangdong Basic and Applied Basic Research Foundation (Grant No.2026A1515012098).

\clearpage
\onecolumngrid

\begin{center}
{\large\bfseries Supplemental Material for ``Topological Word for Non-Abelian Topological Insulators''\par}
\end{center}
\vspace{0.3cm}

\setcounter{section}{0}
\setcounter{equation}{0}
\setcounter{figure}{0}
\setcounter{table}{0}

\renewcommand{\thesection}{S\arabic{section}}
\renewcommand{\theequation}{S\arabic{equation}}
\renewcommand{\thefigure}{S\arabic{figure}}
\renewcommand{\thetable}{S\arabic{table}}

In this Supplemental Material, we provide details on the parameters of the model Hamiltonian in the main text, calculation of the topological invariants, Dirac points, and application of our description to models with long-range hoppings and Floquet systems.

\section{Three-band Model}

In Eq.~(4) of the main text, we consider the three-band tight-binding model of Ref.~\cite{Guo2021}, with three sublattices $A,B,C$ per unit cell. With both nearest-neighbor (NN) and next-nearest-neighbor (NNN) inter-cell hoppings, the Hamiltonian is given by
\begin{equation}
H=\sum_{n}\sum_{X=A,B,C} s_{XX}\,c^\dagger_{X,n}c_{X,n}
+\sum_{n}\sum_{X,Y=A,B,C}\Big(
v_{XY}\,c^\dagger_{X,n}c_{Y,n+1}
+v'_{XY}\,c^\dagger_{X,n}c_{Y,n+2}
+\text{H.c.}\Big).
\label{eq:S_realspace_NN_NNN}
\end{equation}

To make the Bloch Hamiltonian explicitly real under the parity-time (PT) symmetry, we choose purely imaginary inter-cell couplings:
\begin{align}
v_{AB}=v_{BA}=i u,\quad
v_{BC}=v_{CB}=i v,\quad
v_{CA}=v_{AC}=i w,
\label{eq:S_NN_imag}
\\
v'_{AB}=v'_{BA}=i u',\quad
v'_{BC}=v'_{CB}=i v',\quad
v'_{CA}=v'_{AC}=i w',
\label{eq:S_NNN_imag}
\end{align}
with $u,v,w,u',v',w'\in\mathbb{R}$.

The corresponding Bloch Hamiltonian reads
\begin{equation}
H(q)=
\begin{pmatrix}
s_{AA}+2v_{AA}\cos q+2v'_{AA}\cos 2q
& 2u\sin q+2u'\sin 2q
& 2w\sin q+2w'\sin 2q\\
2u\sin q+2u'\sin 2q
& s_{BB}+2v_{BB}\cos q+2v'_{BB}\cos 2q
& 2v\sin q+2v'\sin 2q\\
2w\sin q+2w'\sin 2q
& 2v\sin q+2v'\sin 2q
& s_{CC}+2v_{CC}\cos q+2v'_{CC}\cos 2q
\end{pmatrix}.
\label{eq:S_Bloch_NN_NNN}
\end{equation}
Here $q$ denotes the quasimomentum.

The parameters used in this paper are listed in Tables S1 and S2, with configurations having topological charges \( Q = \pm i, \pm j, \pm k \) in Table S1 and \( Q = -1 \) in Table S2.

\begin{table}[h]
\caption{Flat-band parameter sets for $Q=\pm i,\pm j,\pm k$ ( parameters not listed are zero).}
\begin{ruledtabular}
\begin{tabular}{c c c c c c c c c c}
word & $s_{AA}$ & $s_{BB}$ & $s_{CC}$ & $v_{AA}$ & $v_{BB}$ & $v_{CC}$ & $u$ & $v$ & $w$ \\
\hline
$+i$ & $1$   & $5/2$ & $5/2$ & $0$   & $1/4$  & $-1/4$ & $0$    & $+1/4$ & $0$ \\
$-i$ & $1$   & $5/2$ & $5/2$ & $0$   & $1/4$  & $-1/4$ & $0$    & $-1/4$ & $0$ \\
$+j=ki$ & $2$   & $2$   & $2$   & $1/2$ & $0$    & $-1/2$ & $0$    & $0$    & $-1/2$ \\
$-j=ik$ & $2$   & $2$   & $2$   & $1/2$ & $0$    & $-1/2$ & $0$    & $0$    & $+1/2$ \\
$+k$ & $3/2$ & $3/2$ & $3$   & $1/4$ & $-1/4$ & $0$    & $+1/4$ & $0$    & $0$ \\
$-k$ & $3/2$ & $3/2$ & $3$   & $1/4$ & $-1/4$ & $0$    & $-1/4$ & $0$    & $0$ \\
\end{tabular}
\end{ruledtabular}
\label{tab:S_params_NN}
\end{table}

\begin{table}[h]
\caption{Flat-band parameter set for $Q=-1$ (parameters not listed are zero).}
\begin{ruledtabular}
\begin{tabular}{c c c c c c c c c c}
word & $s_{AA}$ & $s_{BB}$ & $s_{CC}$ & $v'_{AA}$ & $v'_{BB}$ & $v'_{CC}$ & $u'$ & $v'$ & $w'$ \\
\hline
$-1=k^2$ & $3/2$ & $3/2$ & $3$ & $-1/4$ & $+1/4$ & $0$ & $-1/4$ & $0$ & $0$ \\
$-1=i^2$ & $1$   & $5/2$ & $5/2$ & $0$   & $1/4$  & $-1/4$ & $0$    & $+1/4$ & $0$ \\
$-1=kiki$ & $2$   & $2$   & $2$   & $1/2$ & $0$    & $-1/2$ & $0$    & $0$    & $-1/2$ \\
\end{tabular}
\end{ruledtabular}
\label{tab:S_params_NNN}
\end{table}

In the main text, Fig.~2(b) and 2(c) show the spectral interpolation from the trivial phase to the topological phases with topological words $k^2$ and $i^2$, respectively. In fact, such an interpolation is not unique; here we provide details of the interpolation schemes corresponding to Figs.~2(b) and 2(c) in the main text, respectively. We start from the topologically trivial flat-band configuration with all hoppings set to zero, that is,
$v_{XY}=v'_{XY}=0$, and on-site potentials $(s_{AA},s_{BB},s_{CC})=(1,2,3)$.

For the $1\to k^2$ case [Fig.~2(b) in the main text], using the target parameters in Table~\ref{tab:S_params_NNN}, the interpolation reads 
\begin{align}
&s_{AA}=1+\frac{\lambda}{2},\quad
s_{BB}=2-\frac{\lambda}{2},\quad
s_{CC}=3,\\
&v'_{AA}=-\frac{\lambda}{4},\quad
v'_{BB}=+\frac{\lambda}{4},\quad
u'=-\frac{\lambda}{4}.
\end{align}
Parameters not listed above are set to zero for all $\lambda$.

For the $1\to i^2$ case [Fig.~2(c) in the main text], the interpolation is given by
\begin{align}
&s_{AA}=1,\quad
s_{BB}=2+\frac{\lambda}{2},\quad
s_{CC}=3-\frac{\lambda}{2},\\
&v'_{BB}=+\frac{\lambda}{4},\quad
v'_{CC}=-\frac{\lambda}{4},\quad  v'= +\frac{\lambda}{4}.
\end{align}
Again, parameters not listed above are set to zero for all $\lambda$.

In both cases, $\lambda=0$ corresponds to the trivial phase, while $\lambda=1$ reproduces the desired $Q=-1$ configuration.

\section{Calculating topological invariants}

In this section we describe how the quaternion charges are computed from a loop integral of the Wilczek--Berry--Zak (WBZ) connection and explain its $\mathrm{SU}(2)$ lift~\cite{Guo2021, Wu2019}.

For a PT-symmetric three-band system, the Bloch Hamiltonian can be chosen real at all quasimomenta $q$. Consequently, the eigenstates can be chosen as three real orthonormal vectors
$\{|u_1(q)\rangle, |u_2(q)\rangle, |u_3(q)\rangle\}$, forming a moving orthonormal frame in $\mathbb{R}^3$. The order-parameter space is therefore
\begin{equation}
   M_3 = O(3)/O(1)^3 . 
\end{equation}
The non-Abelian WBZ connection is defined as
\begin{equation}
\mathcal{A}_{mn}(q)
= \langle u_m(q) | \partial_q u_n(q) \rangle .
\label{eq:WBZ_def}
\end{equation}

Because the eigenstates are real and orthonormal,
$\partial_q \langle u_m | u_n \rangle = 0$,
we have $\mathcal{A}_{mn} = - \mathcal{A}_{nm}$. Hence, the WBZ connection is antisymmetric, with $\mathcal{A}(q) \in \mathfrak{so}(3)$.
Integrating the connection along a closed loop defines the Wilson loop, which, in the present case, takes values in $SO(3)$.

The connection can be lifted to the double-cover algebra according to
\begin{equation}
\mathcal{A}\in \mathfrak{so}(3) \longrightarrow \widetilde{\mathcal{A}} \in \mathfrak{su}(2).
\end{equation}
This lift follows from the Lie-algebra isomorphism between $\mathfrak{so}(3)$ and $\mathfrak{su}(2)$, under which the generators are mapped as
\begin{equation}
L_x \mapsto \frac{i}{2}\sigma_x, \quad
L_y \mapsto \frac{i}{2}\sigma_y, \quad
L_z \mapsto \frac{i}{2}\sigma_z.
\end{equation}

We therefore define the lifted Wilson loop
\begin{equation}
W[C] = \mathcal{P}\exp\!\left(\oint_C \widetilde{\mathcal{A}}\right),
\label{eq:Wilson}
\end{equation}
where $\widetilde{\mathcal{A}} \in \mathfrak{su}(2)$ is the lifted connection and $\mathcal{P}$ denotes path ordering.

For the PT-symmetric three-band case, the Wilson loop is quantized. 
Since the eigenframe undergoes only $\pi$ or $2\pi$ rotations about the coordinate axes (as discussed in the main text), the Wilson loop reduces to discrete elements of $SU(2)$:
\begin{equation}
W[C] \in \{\pm \sigma_0,\ \pm i\sigma_x,\ \pm i\sigma_y,\ \pm i\sigma_z\}.
\label{eq:SU2_discrete}
\end{equation}

These eight matrices form a representation of the quaternion group $\mathbb{Q}_8$,
with the identification
\begin{equation}
\pm \sigma_0 \leftrightarrow \pm 1, \quad
\pm i\sigma_x \leftrightarrow \pm i, \quad
\pm i\sigma_y \leftrightarrow \pm j, \quad
\pm i\sigma_z \leftrightarrow \pm k .
\end{equation}
Thus the non-Abelian topological charge associated with a loop $C$ is obtained directly from the lifted Wilson loop $W[C]$.

A similar construction applies to the non-Hermitian case when the Hamiltonian still preserves PT symmetry. In this situation, the eigenstates remain real but are no longer orthogonal, so the three eigenvectors no longer form an $O(3)$ frame. To restore a consistent geometric description, one introduces the standard biorthogonal formalism widely used in non-Hermitian band theory~\cite{biorthogonal}. Concretely, we consider the right and left eigenstates $\{|u^R_1(q)\rangle,|u^R_2(q)\rangle,|u^R_3(q)\rangle\}$ and $\{|u^L_1(q)\rangle,|u^L_2(q)\rangle,|u^L_3(q)\rangle\}$, defined through $H(q)|u^R_n(q)\rangle=E_n(q)|u^R_n(q)\rangle$ and $H^\dagger(q)|u^L_n(q)\rangle=E_n^*(q)|u^L_n(q)\rangle$, which satisfy the biorthogonality relation $\langle u^L_m(q)|u^R_n(q)\rangle=\delta_{mn}$. The WBZ connection is then replaced by its biorthogonal form $\mathcal{A}^{\text{LR}}_{mn}(q)=\langle u^L_m(q)|\partial_q u^R_n(q)\rangle$. In the PT-unbroken regime, the spectrum remains real and the resulting Wilson loop constructed from this connection remains quantized, yielding the same discrete $SU(2)$ elements and hence quantized quaternion charges.

\section{Dirac Singularities}
\subsection{A. Path selection and Gauge redundancies}
We first clarify a subtlety while relating interpolation paths to topological words. Two Hamiltonians belong to the same topological phase if there exists a smooth interpolation between them, along which the bulk gap never closes. As a consequence, the topological distinction between two phases cannot be inferred from an arbitrary interpolation, since a given path may contain accidental or redundant singularities that can be removed through smooth deformation. Only those singularities that are unavoidable under any deformation reflect the intrinsic difference between two phases.

Accordingly, to extract the genuine topological information, one should deform the interpolation to a minimal configuration with the smallest number of Dirac points. The resulting word is considered canonical, and characterizes the true topological distinction and the associated physical edge modes. 

A simple example is the interpolation $1\to1$. Since no topological transition occurs, there always exists a fully gapped interpolation with no singularities, corresponding to the empty word $1$. One may nevertheless choose a path that passes through an intermediate phase, for instance, $1\to k\to1$. This necessarily introduces two Dirac points associated with the transitions $1\to k$ and $k\to1$, carrying charges $k$ and $-k$, respectively, each giving rise to a word $k(-k)$. Since the intermediate phase is artificially inserted, this pair of singularities can be removed by a smooth deformation of the path. The resulting word is therefore topologically trivial and does not reflect any genuine distinction in topology, nor does it correspond to protected edge modes.

Having identified a minimal set Dirac points, one can then assign to each a topological charge, and organize them into an ordered word. In this step, however, additional redundancies can arise from the freedom in choosing base points and branch cuts. We now describe the general topological structure underlying this word representation.

Consider a closed loop $C$ in the parameter space that encloses a set of
isolated singularities $\{p_1,\dots,p_N\}$.
After choosing a basepoint $x_0$, the loop can be decomposed into
elementary loops $\gamma_i$ that encircle individual singularities.
According to the van Kampen theorem~\cite{HatcherTopoBook}, these loops generate the fundamental
group of the punctured space, and any loop based at $x_0$ can be written as
an ordered word
$[C]=\gamma_{i_1}\gamma_{i_2}\cdots\gamma_{i_m}$,
where the order of the generators reflects the sequence in which the loop
winds around the singularities.

This word representation is not unique.
Different choices of branch cuts connecting the base point to the
singularities correspond to different loops.
Changing these branch cuts induces local transformations between adjacent
generators, known as Hurwitz moves~\cite{HatcherTopoBook}
\begin{equation}
(\gamma_i,\gamma_{i+1})
\;\longrightarrow\;
(\gamma_{i+1},\,\gamma_{i+1}^{-1}\gamma_i\gamma_{i+1}),
\end{equation}
which permute neighboring singularities along the loop.
Such transformations leave the homotopy class of the loop unchanged,
but modify its word representation.
For example, if two neighboring singularities carry quaternion charges
$(Q_i,Q_{i+1})=(k,i)$, the corresponding word is $j=ki$.
After a Hurwitz move, one obtains $(k,i)\to(i,-k)$, so the word becomes
$i(-k)=-ik$. Hence the topological words $ki$ and $-ki$ are related 
by a gauge transformation and characterize the same phase.

There is also a global redundancy associated with the choice of base point.
Changing the base point amounts to a conjugation of the entire charge
$[C]\rightarrow \gamma_h^{-1}[C]\gamma_h$~\cite{HatcherTopoBook},
which, in terms of total quaternion charge, induces
\begin{equation}
    Q(C)\rightarrow Q_h^{-1}Q(C)Q_h.
\end{equation}
Therefore, the explicit value of the total quaternion element may change
while remaining in the same conjugacy class.
For instance, conjugating $j$ by $i$ gives $i^{-1}ji=-j$,
so shifting the base point can transform the total charge from $j$ to $-j$.

In the physical setting considered here, each elementary loop $\gamma_i$
is associated with a quaternion charge $Q_i\in \mathbb{Q}_8$ obtained from the
lifted Wilson loop.
The total charge carried by the loop $C$ is then $Q(C)=Q_{i_1}Q_{i_2}\cdots Q_{i_m}$,
where the ordered product on the right-hand side is the topological word.
Consequently, the physically meaningful object is the conjugacy class of
the total quaternion element, rather than a particular word representation.
The ordered word keeps track of the detailed sequence of band inversions,
while the global non-Abelian charge characterizes the equivalence class
under these gauge redundancies.

\subsection{B. Higher-Order Singularities and Their Splitting to Dirac Points}

In the main text, we focus on Dirac points as the fundamental type of band degeneracy. 
More generally, other types of singularities may appear. However, we show below that they are non-generic and can always be decomposed into multiple Dirac points under small perturbations.

\begin{figure}[h]
    \centering
    \includegraphics[width=\linewidth]{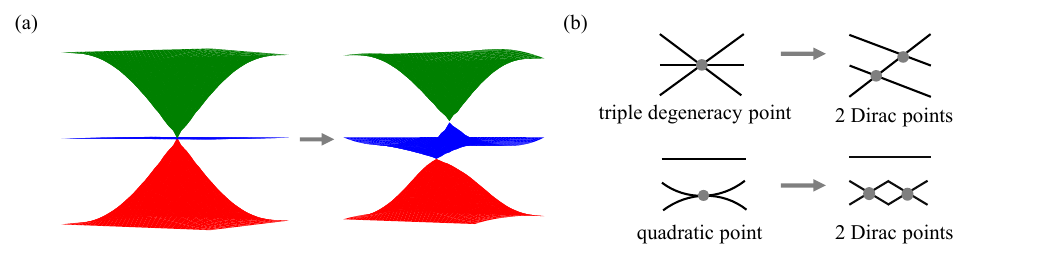}
    \caption{
    Splitting of a higher-order singularity under perturbations.
    (a) Numerical example of the transition from the $1$ phase to the $j=ki$ phase, where a three-fold band degeneracy may appear during the transition (left). By introducing a small perturbation, the triple degeneracy can be split into two Dirac points (right).
    (b) Schematic illustration showing how a non-Dirac singularity can be split into multiple Dirac points under perturbations.
    }
    \label{highorderpoints}
\end{figure}

First, let us consider triple degeneracy points. 
A typical example occurs in the interpolation from the trivial phase to the  $j=ki$ phase. 
Under fine-tuned parameters, two Dirac points may accidentally coincide in momentum space, producing a three-fold band degeneracy. 
Such a configuration is unstable: an arbitrarily small perturbation generically splits the triple degenerate point into two separate Dirac points, as illustrated in Fig.~\ref{highorderpoints}. 
Thus, the triple degeneracy arises only along special parameter sub-manifolds, and does not represent a stable topological defect.

Second, quadratic band-touching points may also occur. 
This situation appears in the phase with an overall charge $Q=-1$, when two Dirac points of the same type overlap. 
Since an isolated Dirac point exhibits linear dispersion, the superposition of two identical Dirac crossings produces a quadratic band touching point. 
Again, this configuration is not topologically protected as a distinct object: a generic perturbation separates it into two Dirac points with linear dispersion.

Such higher-order degeneracies were also noted in Ref.~\cite{Guo2021}.
From the viewpoint of co-dimension analysis, the Dirac point is the generic stable band degeneracy in a one-parameter interpolation of a three-band PT-symmetric system. 
All higher-order singularities require additional fine-tuning and can therefore be resolved into elementary Dirac points.

In summary, although more complicated band-touching structures may appear at special parameter values, they are reducible under perturbations. 
Therefore, Dirac points constitute the universal building blocks of singularities in the present framework.
Constructing word based on Dirac points is therefore 
complete and generically sufficient.

\section{Extension to long-range hoppings and Floquet systems}

\subsection{A. Longer-Range Hoppings and Multiple Edge Modes}
Before discussing explicit examples, we briefly explain the constraint on the word length stated in the main text. Each letter originates from a Dirac point. From the mass-inversion perspective, such a Dirac point corresponds to a domain-wall mode, which, under the open boundary condition, becomes an edge mode. The number of these modes is limited by the boundary capacity, and this capacity is in turn set by the maximal hopping range. Therefore, the number of letters associated with a given gap is bounded by the maximal hopping range. With this in mind, we now consider how longer-range hoppings allow longer words and multiple edge-mode pairs within a given gap.

To describe long-range hoppings, we generalize the real-space Hamiltonian as
\begin{equation}
H=\sum_{n}\sum_{X=A,B,C} s_{XX}\,c^\dagger_{X,n}c_{X,n}
+\sum_{n}\sum_{r=1}^{R}\sum_{X,Y=A,B,C}
\Big(
v^{(r)}_{XY}\,c^\dagger_{X,n}c_{Y,n+r}
+\text{H.c.}
\Big),
\label{eq:S_realspace_longrange}
\end{equation}
where \(R\) denotes the maximum hopping range and \(v^{(r)}_{XY}\) represents the hopping amplitude between unit cells separated by distance \(r\). The nearest- and next-nearest-neighbor hoppings in Eq.~(\ref{eq:S_realspace_NN_NNN}) correspond to the cases with \(r=1\) and \(r=2\), respectively.

To make the Bloch Hamiltonian explicitly real under the parity-time (PT) symmetry, we choose purely imaginary inter-cell couplings for all hopping ranges
\begin{equation}
v^{(r)}_{AB}=v^{(r)}_{BA}=i u_r,\quad
v^{(r)}_{BC}=v^{(r)}_{CB}=i v_r,\quad
v^{(r)}_{CA}=v^{(r)}_{AC}=i w_r,
\label{eq:S_longrange_imag}
\end{equation}
with \(u_r,v_r,w_r\in\mathbb{R}\) for \(r=1,2,\dots,R\).

Take for instance a system with an overall topological charge $-k$. 
Depending on the hopping range, the overall charge can arise from different word representations. 
The simplest case is the elementary word $-k$, which corresponds to a single pair of edge modes.
However, longer-range hopping allows more complicated words with the same overall charge.
More specifically, we can have
\begin{equation}
    -k = k^3 ,
\end{equation}
which corresponds to three pairs of edge modes in the same gap, as illustrated in Fig.~S2(a,b).  
Since the word $k^3$ becomes legal in the presence of intercell hoppings with range greater than three, 
it can be explicitly realized by taking the flat-band parameter set of the $+k$ phase in Table~\ref{tab:S_params_NN} and shifting all nonzero intercell hoppings from $r=1$ to $r=3$. 
Concretely, we choose
$s_{AA}=s_{BB}=3/2$, $s_{CC}=3$, 
$v^{(3)}_{AA}=1/4$, $v^{(3)}_{BB}=-1/4$, and $u_3=1/4$, 
while setting all other hopping amplitudes to zero. 
In the limit where only intercell hoppings of range three are present, the lattice effectively decomposes into three independent chains. 
In the momentum space, this corresponds to traversing three copies of the original Brillouin zone associated with the $k$ phase, giving rise to the word $k^3$ with the total quaternion charge $k^3=-k$.

\begin{figure}[h]
    \centering
    \includegraphics[width=\linewidth]{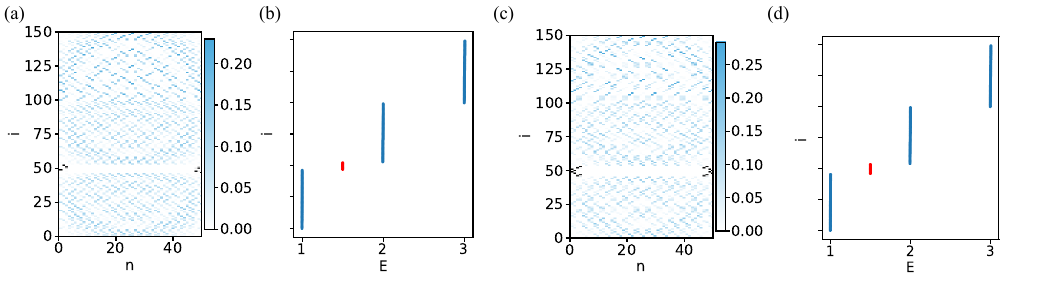}
    \caption{
    (a)(c) Spatial distribution of eigenstates for cases corresponding to the topological words $k^3$ and $k^4$, respectively.
    The horizontal axis denotes the lattice sites, and the vertical axis labels the eigenstates, ordered by their eigenenergies.
    Color bar in blue represents the spatial probability density of the eigenstates, and edge modes are marked in black.    
    (b)(d) Eigenspectra for the corresponding cases $k^3$ and $k^4$.
    The horizontal axis shows the energy and the vertical axis gives the eigenstate index sorted by energy.
    Red dots indicate edge modes. For the calculations shown above, we use the flat-band parameters in which the hopping terms associated with the charge $Q=k$ are replaced by longer-range hoppings, while all other hopping amplitudes are set to zero.
    }
    \label{fig:enter-label}
\end{figure}

An even more striking situation occurs for
\begin{equation}
1 = k^4 .
\end{equation}
Although the total quaternion charge is trivial, the system may still host four pairs of edge modes, 
as shown in Fig.~S2(c,d), which are characterized by the topological word $k^4$. 
For this case, we take the same on-site parameters as the $+k$ phase in Table~\ref{tab:S_params_NN}, while the intercell hoppings are chosen at range $r=4$, namely 
$v^{(4)}_{AA}=1/4$, $v^{(4)}_{BB}=-1/4$, $u_4=1/4$ 
with all other hopping amplitudes set to zero.
This reflects the fact that longer-range hoppings enable the construction of more complicated words.

In general, increasing the hopping range permits longer words to appear and leads to richer edge-state patterns. 
The topological word therefore contains more detailed information than the overall quaternion charge alone: it records the contribution associated with each gap and directly determines the number of in-gap edge modes.
\subsection{B. Floquet Systems}

\begin{figure}[h]
    \centering
    \includegraphics[width=\linewidth]{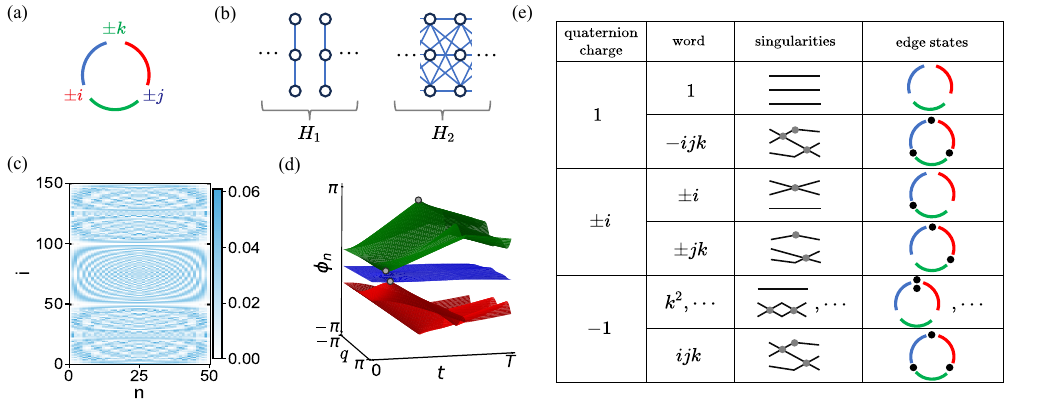}
    \caption{Topological words for Floquet systems.
        (a) Schematic illustration of the quasienergy bands on a circle.
   Each quasienergy gap is associated with a letter, characterizing the nontrivial Zak phase of the band gap.
    (b) Floquet engineering scheme used in the simulation, where the system is driven periodically between two Hamiltonians $H_1$ and $H_2$ within one driving period.
    (c) Spatial distribution of eigenstates obtained by diagonalizing the Floquet operator under the open boundary condition.
    The horizontal axis denotes the lattice sites, while the vertical axis labels the eigenstates ordered by quasienergies.
   The color bar in blue represents the bulk-state probability density, while edge states are marked in black.
      A pair of edge states is visible in each quasienergy gap.
    (d) Phase-band structure in the extended $(q,t)$ manifold showing the phase-band singularities.
    Three Dirac points appear, corresponding to the generators of the topological word.
    (e) Summary of the correspondence between the quaternion charge, word representation, phase-band singularities, and the distribution of edge modes across different quasienergy gaps.
    }
    \label{fig:floquet}
\end{figure}

We now extend the word formalism to periodically driven Floquet systems.
Therein, topology is defined for the quasienergy bands,
which are the eigen phase factors of the Floquet operator $U$. 
Since quasienergies are defined modulo $2\pi/T$ ($T$ is the period of the Floquet drive), the quasienergy spectrum lives on a closed loop rather than on $\mathbb R$, as illustrated in Fig.~\ref{fig:floquet}(a).
Consequently, in addition to the conventional band-adjacency structure, the highest and lowest quasienergy bands are also adjacent.
This additional Floquet $\pi$ gap introduces an extra generator in the letter set.
In particular, in a three-band system, besides the static generators $\pm i$ and $\pm k$,
the quasienergy bands legalize the generator $\pm j$.
The full set of elementary generators in the Floquet setting is therefore
\begin{equation}
\mathcal{E}=\{\pm i,\pm j,\pm k\}.
\end{equation}

As an explicit example, we follow the Floquet engineering scheme in Ref.~\cite{Li2023}, where the Floquet operator is given by
\begin{equation}
U(T)=e^{-iH_2T/2}e^{-iH_1T/2},
\end{equation}
where
\begin{align}
H_1=\sum_{n=1}^{L}\sum_{X,Y}
s_{XY}\,c^\dagger_{X,n}c_{Y,n}
+\mathrm{H.c.},\\
H_2=\sum_{n=1}^{L-1}\sum_{X,Y}
v_{XY}\,c^\dagger_{X,n}c_{Y,n+1}
+\mathrm{H.c.}.
\end{align}
Here $c_{X,n}$ annihilates a particle on sublattice $X\in\{A,B,C\}$
in the $n$th unit cell. As illustrated in Fig.~\ref{fig:floquet}(b), 
$H_1$ activates intra-cell hoppings for the first half period, and 
$H_2$ activates inter-cell hoppings for the second half period.

As discussed in Ref.~\cite{Li2023}, one observes that the same quaternion charge may correspond to different edge-mode configurations.
For example, the phase with an overall topological charge $1$ can exhibit edge modes in all three gaps,
as shown in Fig.~\ref{fig:floquet}(c). 
This occurs because the trivial quaternion charge does not imply
a unique word representation:
in addition to the empty word $1$,
nontrivial products such as 
\begin{equation}
1=kji,
\end{equation}
are also allowed. 
Thus, the same overall charge may correspond to distinct gap-resolved structures.

To make this structure explicit,
one may construct a continuous interpolation between a Floquet topological phase
and the trivial phase.
In Floquet systems, this interpolation naturally arises
by introducing the time parameter $t\in[0,T]$
and considering the evolution operator $U(t)$.
At $t=0$ one has $U(0)=1$,
which is topologically trivial and corresponds to the empty word.
As $t$ increases to $T$,
singularities in the phase bands may appear and disappear.
Following the procedure described in Ref.~\cite{Li2023},
the phase bands can be obtained from a continuous deformation
of the evolution operator, revealing the associated singularities
in the $(q,t)$ space, as shown in Fig.~\ref{fig:floquet}(d). The specific parameter choices can be found in Ref.~\cite{Li2023}.

From the perspective of topological word, the phase-band construction is equivalent to building a continuous mapping between a nontrivial phase and the trivial phase. Singularities encountered along this interpolation precisely encode the construction of topological word.
Our framework thus provides a unified description for the non-Abelian bulk-boundary correspondence in both static and Floquet systems. 

The complete correspondence among words, singularities, and edge-mode multiplicities is summarized in Fig.~\ref{fig:floquet}(e).
Notably, Floquet quasienergies correspond to the eigenphases of the evolution operator, and the resulting effective Floquet Hamiltonian generally contains longer-range hoppings. Consequently, more complex word structures and edge-state patterns can arise. 
For the convenience of discussion, here we restrict our attention to cases where simple topological words are sufficient to characterize the non-Abelian bulk-boundary correspondence.

\end{document}